\documentclass[12pt,preprint]{aastex}

\shorttitle{Emission from a Black Hole}
\shortauthors{Fujita}

\begin{document}

\title{X-Ray Emission from a Supermassive Black Hole Ejected from the
Center of a Galaxy}

\author{Yutaka Fujita}
\affil{Department of Earth and Space Science, Graduate School
of Science, Osaka University, \\
1-1 Machikaneyama-cho, Toyonaka, Osaka
560-0043, Japan}

\begin{abstract}
 Recent studies have indicated that the emission of gravitational waves
 at the merger of two black holes gives a kick to the final black
 hole. If the supermassive black hole at the center of a disk galaxy is
 kicked but the velocity is not large enough to escape from the host
 galaxy, it will fall back onto the the disk and accrete the
 interstellar medium in the disk. We study the X-ray emission from the
 black holes with masses of $\sim 10^7\: M_\sun$ recoiled from the
 galactic center with velocities of $\sim 600\rm\: km\: s^{-1}$. We find
 that their luminosities can reach $\ga 10^{39}\rm\: erg\: s^{-1}$, when
 they pass the apastrons in the disk. While the X-ray luminosities are
 comparable to those of ultra-luminous X-ray sources (ULXs) observed in
 disk galaxies, ULXs observed so far do not seem to be such supermassive
 black holes. Statical studies could constrain the probability of merger
 and recoil of supermassive black holes.
\end{abstract}


\section{Introduction}

Recently, studies in numerical general relativity have shown that merged
binary black holes can have a large recoil velocity through anisotropic
emission of gravitational waves \citep[e.g.][]{gon07,cam07}. The maximum
velocity would reach $\sim 4000\rm\: km\: s^{-1}$, although the actual
distribution of kick velocities is very uncertain. The discovery of such
recoiled black holes is important for studies about the growth of black
holes as well as the general relativity.

Supermassive black holes ($\ga 10^6\: M_\sun$) at the centers of disk
galaxies would be kicked through this mechanism. Although they would be
bright just after the kick because of the emission from the accretion
disk carried by the black holes, they would soon get dim as the disk is
consumed by the black holes \citep{loe07}. However, if the kick velocity
of a black hole is not large enough to escape from the host galaxy, it
will eventually fall back onto the galactic disk. If the time-scale of
dynamical friction is large enough, it will revolve around the galactic
center many times before it finally spirals into the galactic
center. When the black hole passes the galactic disk, it will accrete
the interstellar medium (ISM) in the disk \citep{ble08}.

The accretion of the surrounding gas onto an isolated black hole has
been studied by several authors \citep*[][and references
therein]{fuj98,ago02,mii05,map06}. Most of the previous studies focused
on stellar mass ($\sim 10\: M_\sun$) or intermediate mass black holes
(IMBHs; $\sim 10^3\: M_\sun$). The accretion rate and thus the
luminosity of a black hole depend on the mass of the black hole and the
density of the gas surrounding it (see equation~[\ref{eq:dotm}]). Since
the mass of a stellar-mass black hole is small, its luminosity becomes
large enough to be observed only when it plunges into a high density
region such as a molecular cloud. On the other hand, in this letter, we
show that a recoiled supermassive black hole can shine even in the
ordinary region of a galactic disk because of its huge mass.

\section{Models}

We calculate the orbit of a recoiled black hole in a fixed galaxy
potential. The galaxy potential consists of three components, which are
\citet{miy75} disk, Hernquist spheroid, and a logarithmic halo:
\begin{equation}
 \Phi_{\rm disk}=-\frac{G M_{\rm dist}}
{\sqrt{R^2+(a+\sqrt{z^2+b^2})^2}}\:,
\end{equation}
\begin{equation}
\Phi_{\rm sphere}=-\frac{G M_{\rm sphere}}{r+c}\:,
\end{equation}
\begin{equation}
\Phi_{\rm halo}=\frac{1}{2}v_{\rm halo}^2
\ln\left[R^2+\left(\frac{z}{q}\right)^2+d^2\right]\:,
\end{equation}
where $R$ ($=\sqrt{x^2+y^2}$) and $z$ are cylindrical coordinates
aligned with the galactic disk, and $r=\sqrt{R^2+z^2}$. We adopt the
parameters for the Galaxy. We take $M_{\rm disk}=1.0\times 10^{11}\:
M_\sun$, $M_{\rm sphere}=3.4\times 10^{10}\: M_\sun$, $a=6.5$~kpc,
$b=0.26$~kpc, $c=0.7$~kpc, $d=13$~kpc, and $q=0.9$; $v_{\rm halo}$ is
determined so that the circular velocity for the total potential is
$220\rm\: km\: s^{-1}$ at $R=7$~kpc \citep[see][]{law05}. We solve the
equation of motion for the supermassive black hole:
\begin{equation}
\label{eq:motion}
 \mbox{\boldmath{$\dot{v}$}}=-\nabla\Phi\:,
\end{equation}
where {\boldmath{$v$}}$=(v_x, v_y, v_z)$ is the velocity of the black
hole, and $\Phi=\Phi_{\rm disk}+\Phi_{\rm sphere}+\Phi_{\rm halo}$. The
density of the disk is given by
\begin{equation}
 \rho_{\rm disk}=\left(\frac{b^2 M_{\rm disk}}{4\pi}\right)
\frac{a R^2+(a+3\sqrt{z^2+b^2})(a+\sqrt{z^2+b^2})^2}
{[R^2+(a+\sqrt{z^2+b^2})^2]^{5/2}(z^2+b^2)^{3/2}}
\end{equation}
\citep{miy75}. We assume that part of the disk consists of the ISM; its
density is represented by $\rho_{\rm ISM}=f_{\rm ISM}\rho_{\rm disk}$
and $f_{\rm ISM}=0.2$. The circulation velocity of the disk is given by
\begin{equation}
 v_{\rm cir}=\sqrt{r\frac{\partial \Phi}{\partial r}} \:.
\end{equation}

The accretion rate of the ISM onto the supermassive black hole is given by
the Bondi-Hoyle accretion \citep{bon52}:
\begin{equation}
\label{eq:dotm}
 \dot{m}=2.5\pi G^2
\frac{m_{\rm BH}^2\rho_{\rm ISM}}{(c_s^2+v_{\rm rel}^2)^{3/2}}\:,
\end{equation}
where $m_{\rm BH}$ is the mass of the black hole, $c_s$ ($=10\rm\: km\:
s^{-1}$) is the sound velocity of the ISM, and $v_{\rm rel}$ is the
relative velocity between the black hole and the surrounding ISM. We
assume that the orbit of the black hole is confined on the $x$-$z$ plane
($v_y=0$). Thus, the relative velocity is simply given by $v_{\rm
rel}^2=v_x^2+v_{\rm cir}^2+v_z^2$.
The X-ray luminosity of the black hole is given by
\begin{equation}
 L_{\rm X}=\eta \dot{m}c^2\:,
\end{equation}
where $\eta$ is the efficiency.  Since the accretion rate is relatively
small for the mass of the black hole, the accretion flow would be a
radiatively inefficient accretion flow \citep[RIAF;][]{ich77,nar95}. In
this case, the efficiency follows $\eta\propto\dot{m}$ for $L_{\rm X}\la
0.1 L_{\rm Edd}$, where $L_{\rm Edd}$ is the Eddington luminosity
\citep*[e.g.][]{kat98}. Therefore, we assume that $\eta=\eta_{\rm Edd}$
for $\dot{m}> 0.1 \dot{m}_{\rm Edd}$ and $\eta = \eta_{\rm
Edd}\dot{m}/(0.1\dot{m}_{\rm Edd})$ for $\dot{m}< 0.1 \dot{m}_{\rm
Edd}$, where $\dot{m}_{\rm Edd}=L_{\rm Edd}/(c^2 \eta_{\rm Edd})$
\citep{mii05}. We assume that $\eta_{\rm Edd}=0.1$.

We solved equation~(\ref{eq:motion}) by Mathematica 6.0 using a command
NDSolve. The black hole is ejected on the $x$-$z$ plane at $t=0$. The
direction of the ejection changes from $\theta=0\arcdeg$ to $90\arcdeg$,
where $\theta=0\arcdeg$ corresponds to the $z$-axis. We calculate the
orbit until $t=t_{\rm max}$, which is chosen to be much larger than the
period of revolution and to be smaller than the time-scale of dynamical
friction. The latter is estimated to be
\begin{equation}
\label{eq:dynfrc}
 t_{\rm df}=\frac{v_{\rm rel}}{\dot{v}_{\rm rel}}
=\frac{v_{\rm rel}^3}{4\pi G^2 m_{\rm BH}
\rho\ln\Lambda}\:
\end{equation}
\citep{bin08}, where $\rho$ is the total density
(disk$+$sphere$+$halo). The halo component does not much affect the
dynamical friction. Since $N$-body simulations for a spherically
symmetric potential showed that the Coulomb logarithm is $\ln\Lambda\sim
2$--3 \citep{gua08}, we take $\ln\Lambda=2.5$. The effects of dynamical
friction on orbits in a complex potential like the one we adopted would
be complicated and ideally should be studied with high-resolution
$N$-body simulations. Thus, equation~(\ref{eq:dynfrc}) should be
regarded as a rough estimate of the time-scale of the
dynamical-friction.

\section{Results}
\label{sec:results}

The black hole is placed at the center of the galaxy at $t=0$. Since we
do not know the distributions of mass and initial velocity ($v_0$) of
the black hole, we consider situations in which the emission from it
would be observed easily. That is, the luminosity of the black hole
would be large, and the observable time would be long.

We consider five combinations of $m_{\rm BH}$ and $v_0$ shown in
Table~\ref{tab:par}. If we take larger $m_{\rm BH}$ and/or smaller
$v_0$, the dynamical friction becomes more effective and the black hole
quickly falls into the galaxy center. On the other hand, if we take
smaller $m_{\rm BH}$ and/or larger $v_0$, the luminosity of the black
hole becomes too small to be observed
(equation~[\ref{eq:dotm}]). Moreover, the black hole is not bound to the
galaxy, if $v_0$ is too large. The dynamical friction is most effective
when $\theta=90\arcdeg$. In Table~\ref{tab:par}, we show the
time-average of the time-scale, $\langle t_{\rm
df}\rangle_{\theta=90\arcdeg}$, for $0<t<t_{\rm max}$ and
$\theta=90\arcdeg$.

\begin{figure}
\epsscale{.80} \plotone{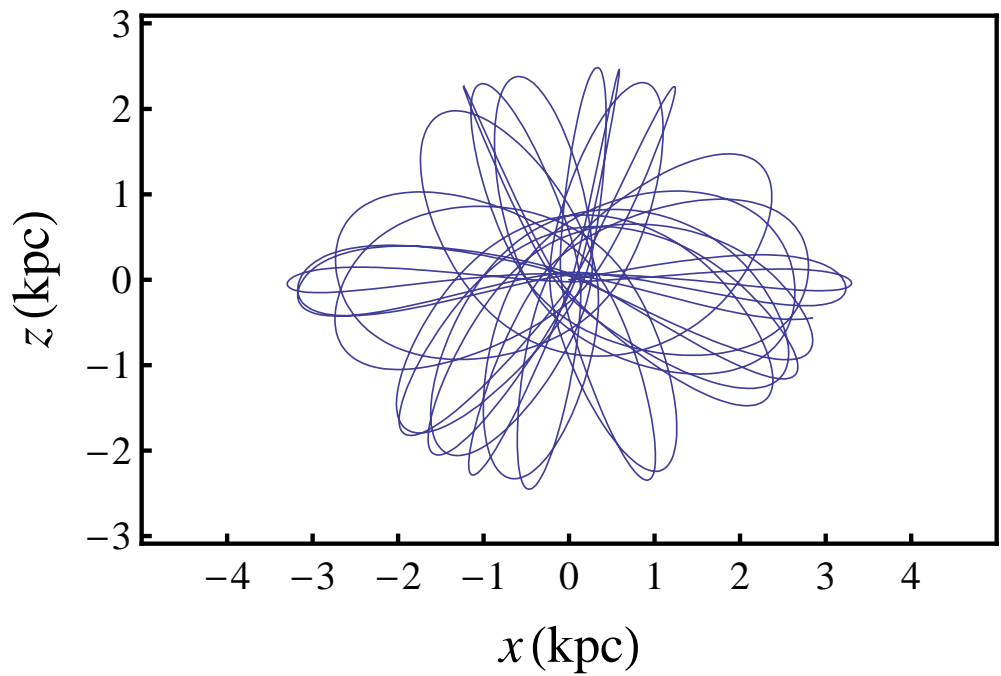} \caption{The orbit of a black hole for
$0<t<1$~Gyr when $v_0=600\rm\: km\: s^{-1}$ and
$\theta=80\arcdeg$. \label{fig:orbit}}
\end{figure}

\begin{figure}
\epsscale{.80} \plotone{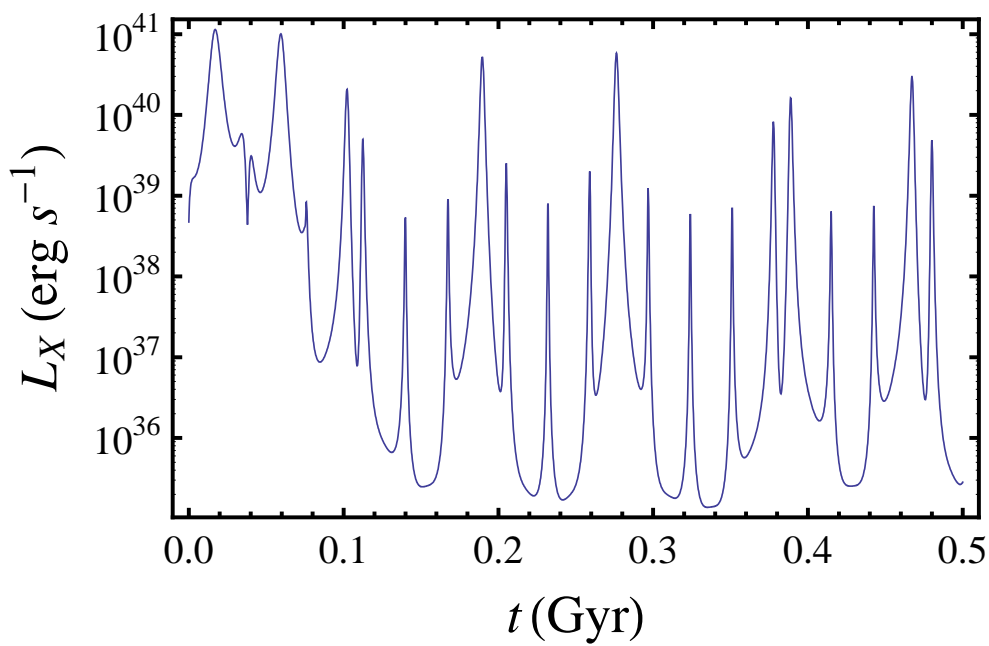} \caption{The luminosity of a black
hole for $0<t<0.5$~Gyr when $m_{\rm BH}=3\times 10^7\: M_\sun$,
$v_0=600\rm\: km\: s^{-1}$ and $\theta=80\arcdeg$. \label{fig:lum}}
\end{figure}

Fig.~\ref{fig:orbit} shows the orbit of the black hole when
$v_0=600\rm\: km\: s^{-1}$ and $\theta=80\arcdeg$. Fig.~\ref{fig:lum}
shows the luminosity of the same black hole ($m_{\rm BH}=3\times 10^7\:
M_\sun$). In Table~\ref{tab:par}, we present the distance of the
apastrons from the center of the galaxy ($r_{\rm max}$) when
$\theta=90\arcdeg$. It is to be noted that $r_{\rm max}$ is not much
dependent on $\theta$ for a given $v_0$. We also present the maximum
X-ray luminosity of the black hole ($L_{\rm max}$) when
$\theta=90\arcdeg$ in Table~\ref{tab:par}. For a given $m_{\rm BH}$ and
$v_0$, the X-ray luminosity is larger when $\theta$ is closer to
$90\arcdeg$, because the orbit is included in the galactic disk, where
$\rho_{\rm ISM}$ is large.

We found that for $v_0\la 600\rm\: km\: s^{-1}$, the luminosity reaches
its maximum when the black hole passes apastrons and when the apastrons
reside in the disk of the galaxy. This is because $v$ decreases,
$\rho_{\rm ISM}$ increases, and thus $\dot{m}$ increases there
(equation~[\ref{eq:dotm}]). On the other hand, for $v_0\sim 700\rm\:
km\: s^{-1}$, the distance of the apastrons from the galactic center
($r_{\rm max}$) is always large (Table~\ref{tab:par}). Thus, even if
apastrons reside in the disk, $\rho_{\rm ISM}$ is small
there. Therefore, the luminosity of the black hole reaches its maximum
between the apastron and periastron, and $L_{\rm max}$ is smaller
compared with the models of $v_0\la 600\rm\: km\: s^{-1}$
(Table~\ref{tab:par}).

\begin{figure}
\epsscale{.80} \plotone{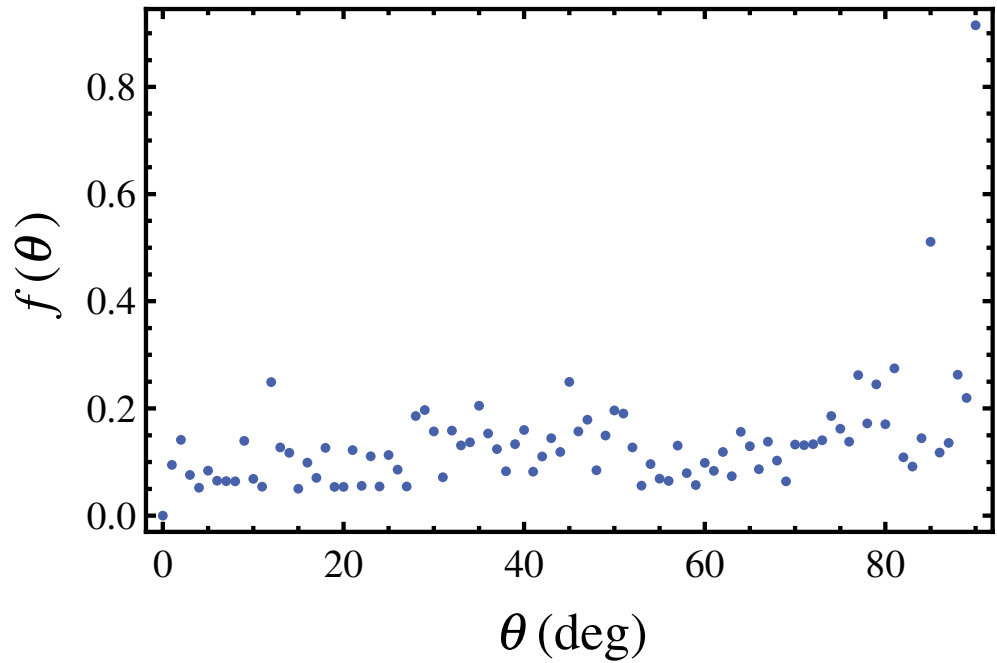} \caption{The fraction of the period
during which the black hole luminosity becomes larger than $L_{\rm th}$,
when $m_{\rm BH}=3\times 10^7\: M_\sun$, $v_0=600\rm\: km\: s^{-1}$, and
$L_{\rm th}=3\times 10^{39}\rm\: erg\: s^{-1}$. \label{fig:ftheta}}
\end{figure}

Assuming that black holes are ejected in random directions at the
centers of galaxies, we estimate the probability of observing black
holes with luminosities larger than a threshold luminosity $L_{\rm
th}$. For given $m_{\rm BH}$ and $v_0$, we calculate 91 evolutions of
the luminosity by changing $\theta$ from $0\arcdeg$ to $90\arcdeg$ by
one degree. Then, we obtain the period during which the relation $L_{\rm
X}>L_{\rm th}$ is satisfied for each $\theta$, and divide the period by
$t_{\rm max}$. This is the fraction of the period during which the black
hole luminosity becomes larger than $L_{\rm th}$. We refer to this
fraction as $f(\theta)$ and show it in Fig.~\ref{fig:ftheta} when
$m_{\rm BH}=3\times 10^7\: M_\sun$, $v_0=600\rm\: km\: s^{-1}$, and
$L_{\rm th}=3\times 10^{39}\rm\: erg\: s^{-1}$. We average $f(\theta)$
by $\theta$, weighting with $\sin\theta$, and obtain the probability of
observing black holes with $L_{\rm X}>L_{\rm th}$. In
Table~\ref{tab:par}, we present the probability $P_{3e39}$ when $L_{\rm
th}=3\times 10^{39}\rm\: erg\: s^{-1}$; for the parameters we chose,
$P_{3e39}\sim 0$--0.56.

\section{Discussion}

We have found that a supermassive back hole that had been recoiled at
the center of a disk galaxy could be observed in the galactic disk with
an X-ray luminosity of $L_{\rm X}\ga 10^{39}\rm\: erg\: s^{-1}$. One of
the candidates of such objects is ultraluminous X-ray sources (ULXs)
observed in disk galaxies \citep{col99,mak00,mus04}. They are found in
off-nuclear regions of nearby galaxies and their X-ray luminosities
exceed $\sim 3\times 10^{39}\rm\: erg\: s^{-1}$, which are larger than
the Eddington luminosity of a black hole with a mass of $\sim 20\:
M_\sun$. If ULXs are stellar mass black holes, they might be explained
by anisotropic emission \citep{rey97,kin01}, slim-disks \citep*{wat01}
or thin, super-Eddington accretion disks \citep{beg02}. On the other
hand, there is some evidence that they are IMBHs at least for some of
them \citep*{mil04,cro04}.

Considering their X-ray luminosities and off-center positions, some of
the ULXs might be the recoiled supermassive black holes. However, the
fraction of supermassive black holes in the ULXs would not be
large. \citet{sch07} estimated that for comparable mass binaries with
dimensionless spin values of 0.9, only $\sim 10$\% of all mergers are
expected to result in an ejection speed of $\sim 500$--$700\rm\: km\:
s^{-1}$. Since the ejection speed is smaller for mergers with large mass
ratios and smaller spin values, the actual fraction would be
smaller. Moreover, in our model, the time-corrected probability of
observing black holes with $L_{\rm X}>3\times 10^{39}\rm\: erg\: s^{-1}$
is $\tilde{P}_{3e39}\la 0.1$, where $\tilde{P}_{3e39}$ is obtained by
averaging $\min[\langle t_{\rm df}\rangle_{\theta},t_{\rm age}]
f(\theta)/t_{\rm age}$ by $\theta$, weighting with $\sin\theta$, and
$t_{\rm age}$ ($\sim 10$~Gyr) is the age of a galaxy
(Table~\ref{tab:par}). Here, we note that $\langle t_{\rm
df}\rangle_{\theta}$ should be regarded as the upper-limit of the actual
time-scale, because $t_{\rm df}$ should decrease through the dynamical
friction every time the black hole passes the dense region of the
galaxy. Furthermore, our model indicates that a traveling supermassive
black hole needs to have a mass comparable to the one currently observed
at the galactic center in order to have large $L_X$. It is unlikely that
a galaxy would have undergone many mergers of black holes with such
masses. The number of such mergers that a galaxy has undergone would be
$N\la 1$. Thus, the probability that a galaxy has a traveling
supermassive black hole with a luminosity comparable to that of ULXs is
$\la 1\times 10^{-2}$.

In fact, current radio observations seem to show that ULXs observed so
far are not supermassive black holes. Our model predicts that the X-ray
luminosity of a supermassive black hole traveling through the galaxy is
comparable to the typical X-ray luminosity of a LINER \citep[$\sim
4\times 10^{39}$--$5\times 10^{41}\:\rm erg\: s^{-1}$;][]{ter02}. LINERs
seem to show core radio emission and many even have detectable jets
\citep{nag05}. On the other hand, radio observations have shown that no
ULX has been detected with a unresolved radio core
\citep{mus04}. Moreover, it has been shown that the optical luminosities
of ULXs tend to be smaller than their X-ray luminosities
\citep[e.g.][]{pta06}, which is inconsistent with typical RIAF spectra
\citep*[e.g.][]{yua04}. Thus, it is unlikely that most of the ULXs are
the supermassive black holes traveling through the galaxies.

However, the recoiled supermassive black holes could be found through
future extensive surveys. Our model predicts that the X-ray luminous
black holes should not be observed far from the centers of the host
galaxies (say $\ga 10$~kpc), because $\rho_{\rm ISM}$ should be small
there (\S~\ref{sec:results}). Our model also predicts that the relative
velocity between the X-ray source and the surrounding ISM and stars is
$v_{\rm rel}\ga v_{\rm cir}$. If atomic line emission associated with
the X-ray source is observed, the velocity could be estimated through
the Doppler shift. Instead of X-rays, \citet{mac05} argued that radio
detections may be best to search for isolated accreting black holes. The
detailed analysis of the spectra and the time variability would be
useful to determine the masses of the black holes \citep{mus04}. In the
future, statistical studies could observationally constrain the
probability of the mergers of black holes and the recoil.

\section{Conclusion}

We have shown that a supermassive black hole ejected from the center of
the host disk galaxy will return to the galactic disk, if the initial
velocity is smaller than the escape velocity of the galaxy. The black
hole accretes the surrounding ISM and the resultant X-ray luminosity can
reach $\ga 10^{39}\:\rm erg\: s^{-1}$, when it passes the apastrons in
the disk. Although the luminosity of a recoiled supermassive black hole
is comparable to that of ultra-luminous X-ray sources (ULXs), it is
unlikely that many of the observed ULXs are the supermassive black
holes.

\acknowledgments

I would like to thank the anonymous referee for useful comments. I am
grateful to H.~Tagoshi and T.~Tsuribe for useful discussion. YF was
supported in part by Grants-in-Aid from the Ministry of Education,
Culture, Sports, Science and Technology of Japan (20540269).

\clearpage

\begin{deluxetable}{cccccccc}
\tablewidth{0pt}
\tablecaption{Model Parameters and Results\label{tab:par}}
\tablehead{
\colhead{$m_{\rm BH}$} &
\colhead{$v_0$} &
\colhead{$t_{\rm max}$} &
\colhead{$\langle t_{\rm df}\rangle_{\theta=90\arcdeg}$} &
\colhead{$r_{\rm max}$} &
\colhead{$L_{\rm max}$} &
\colhead{$P_{\rm 3e39}$} &
\colhead{$\tilde{P}_{\rm 3e39}$} \\
\colhead{($M_\sun$)} &
\colhead{($\rm km\: s^{-1}$)} &
\colhead{(Gyr)} &
\colhead{(Gyr)} &
\colhead{(kpc)} &
\colhead{($\rm erg\: s^{-1}$)} &
\colhead{} &
\colhead{} 
}
\startdata
$3\times 10^6$&500& 0.3 &  4.7 &  1 & $1\times 10^{38}$ & 0     & 0    \\
$1\times 10^7$&500& 0.3 &  1.4 &  1 & $5\times 10^{39}$ & 0.064 & 0.010\\
$1\times 10^7$&600&   1 &  4.5 &  3 & $5\times 10^{39}$ & 0.016 & 0.012\\
$1\times 10^7$&700&   2 &   27 & 13 & $8\times 10^{37}$ & 0     & 0    \\
$3\times 10^7$&500& 0.3 & 0.47 &  1 & $1\times 10^{41}$ & 0.56  & 0.031\\
$3\times 10^7$&600&   1 &  1.5 &  3 & $1\times 10^{41}$ & 0.15  & 0.11 \\
$3\times 10^7$&700&   2 &  9.0 & 13 & $2\times 10^{39}$ & 0     & 0    \\
\enddata
\end{deluxetable}

\clearpage

\end{document}